\documentclass[english,12pt]{article}
          \usepackage{babel}
          \usepackage[T1]{fontenc}
          \usepackage[latin2]{inputenc}
          \usepackage{pslatex}
\begin{document}


\title{First results from the use of the relativistic and slim disc model SLIMULX in XSPEC}

\author{
       M. D. Caballero-Garcia$^1$, M. Bursa$^1$, M. Dov{\v c}iak$^1$, S. Fabrika$^2$, \\
       A. J. Castro-Tirado$^3$ and V. Karas$^1$
       }

\date{
         $1$Astronomical Institute of the Academy of Sciences, 
             Bo{\v c}ni 1401, CZ-14100 Praha, Czech Republic, email: garcia@asu.cas.cz \\ 
         $2$Special Astrophysical Observatory, Nizhnij Arkhyz 369167, Russia \\
         $3$Instituto de Astrof\'{\i}sica de Andaluc\'{\i}a (IAA-CSIC), P.O. Box 03004, E-18080, Granada, Spain 
          }

\maketitle

\begin{abstract}

Ultra-Luminous X-ray sources (ULXs) are accreting black holes for which their X-ray 
properties have been seen to be different to the case of stellar-mass black hole binaries. For 
most of the cases their intrinsic energy spectra are well described by a cold accretion disc 
(thermal) plus a curved high-energy emission components. The mass of the black hole (BH) derived 
from the thermal disc component is usually in the range of 100-1000 solar masses, which have 
led to the idea that this might represent strong evidence of the Intermediate Mass Black Holes 
(IMBH), proposed to exist by theoretical studies but with no firm detection (as a class) so 
far. Recent theoretical and observational developments are leading towards the idea that these 
sources are instead stellar-mass BHs accreting at an unusual super-Eddington regime. In this paper
we briefly describe the model SLIMULX that can be used in XSPEC for the fit 
of thermal spectra of slim discs around stellar mass black holes in the super-Eddington regime. This model
consistently takes all relativistic effects into account. We present the obtained results 
from the fit of the X-ray spectra from NGC~5408 X--1.

\end{abstract}

\section{Introduction}

Ultra-Luminous X-ray sources (ULXs) are point-like, off-nuclear, extra-galactic sources, with observed X-ray luminosities
(${\rm L}_{\rm X}{\ge}10^{39}\,{\rm erg}\,{\rm s}^{-1}$) higher than the Eddington luminosity for a stellar-mass black-hole
(${\rm L}_{\rm X}{\approx}10^{38}\,{\rm erg}\,{\rm s}^{-1}$). The true nature of these objects is still debated
(Feng \& Soria, 2011; Fender \& Belloni, 2012) as there is still no unambiguous estimate for the mass of the compact object hosted in these systems.

Assuming an isotropic emission, in order to avoid the violation of the Eddington limit, ULXs
might be powered by accretion onto Intermediate Mass Black Holes (IMBHs) with masses in the range
$10^{2}-10^{5}\,{\rm M}_{\odot}$ (Colbert \& Mushotzky, 1999; Greene \& Ho, 2007; Farrell et al., 2009). Recently, some 
studies have shown evidence of some ULXs being Black Hole Binaries (e.g. M~82 X--2; Kong et al., 2007; Caballero-Garcia et al., 2013a). Later 
it was shown that M~82 X--2 is a binary but
accreting onto a neutron star (Bachetti et al., 2014). It
is suggested that ULXs are also powered by accretion onto stellar-mass black holes $(<100\,{\rm M}_{\odot})$ at around or in
excess of the Eddington limit (e.g. 
Colbert \& Mushotzky, 1999; Fabrika \& Mescheryakov, 2001; King et al., 2001; Fabbiano, 2006, Poutanen et al. 2007; Liu et al. 2013).


Initially, they were supposed to be the IMBHs originating from low-metallicity Population III stars (Madau \& Rees, 2001). Nevertheless, 
they are not spatially distributed throughout galaxies as it would be expected. On the other hand, 
IMBHs may be produced in runaway mergers in the cores of young clusters (Portegies Zwart, 2004). In such cases, they usually  
stay within their clusters. It has been found (Poutanen et al., 2013) that all brightest X-ray sources in the Antennae galaxies are located 
nearby the very young stellar clusters. NGC~5408 X--1 is also located nearby a young stellar association (Gris\'e et al., 2012). These studies 
concluded that these sources were ejected in the process of formation of stellar clusters in the dynamical few-body encounters and that the 
majority of ULXs are massive X-ray binaries with the progenitor masses larger than $50\,{\rm M}_{\odot}$. Currently, it is thought that 
only a handful of ULXs could be considered as potential IMBHs (ESO~243-49 HLX-1 between a few others; Farrell et al., 2009; Sutton et al. 2012).

In this paper we analyze one of the best available spectra from the ULX NGC~5408 X--1 obtained so far including the model {\tt SLIMULX} developed for fitting of
thermal spectra of slim-discs around stellar mass black holes in the super-Eddington regime. We first introduce important observational results
obtained for NGC~5408 X--1 so far in Sec.~1.1. Afterwards we give brief descriptions of the accretion disc theory both using standard
and slim-disc configurations (Sec.~2), in order to justify the use of the latter in the case of NGC~5408 X--1. Finally, we present and 
discuss the analysis and results obtained by using {\tt SLIMULX} (Sec.~3).

\subsection{NGC~5408 X--1} \label{ngc5408}

The ULX in NGC~5408 X--1 was discovered with the {\it Einstein} observatory (Stewart et al. 1982) and its {\it soft excess} found with {\it ROSAT}
(Fabian \& Ward, 1993). It is located in a close-by (${\rm D}=4.8$\,Mpc, Karachentsev et al., 2003) small (size of $2.2{\times}1.1$\,kpc) star-burst galaxy 
and at ${\approx}20$\,arcsec from its centre. NGC~5408 X--1 is very bright, with a peak X-ray luminosity in the (0.3-10\,keV)
energy range of ${\rm L}_{\rm X}=2{\times}10^{40}\,{\rm erg}\,{\rm s}^{-1}$. Strohmayer \& Mushotzky (2009)
found a QPO in its PDS centred at $0.01$\,Hz and inferred a mass for the black hole in the range of $10^{3}-10^{4}\,{\rm M}_{\odot}$. On the other hand, Middleton et al. (2011) proposed
a much smaller mass ($10^2\,{\rm M}_{\odot}$) in base of the QPO and the timing properties. They proposed that NGC~5408 X--1 is accreting in a super-Eddington regime and that
the QPO is analogous to the ultra-Low-Frequency QPO seen occasionally in a few black hole binaries (BHBs). More recently, Dheeraj \& Strohmayer (2012) studied the timing and
spectral properties of NGC~5408 X--1 and have found that the QPO frequency is variable (within the frequency range of 0.0001-0.19\,Hz) and that the spectral properties are approximately 
constant. They suggested that NGC~5408 X--1 is accreting in the {\it saturation regime} (increase of the QPO frequency with constant disc flux and power-law photon index) 
frequently observed in BHBs (Vignarca et al., 2003). On the other hand, based on the comparison with the case of BHBs but in the framework of the accretion states around 
BHs Caballero-Garcia et al. (2013b) studied its spectral-timing properties during the same time-lapse as done by Dheeraj \& Strohmayer (2012) (i.e. 6\,yrs) and found a close 
similarity with the hard-intermediate state of BHBs but accreting at a much higher luminosity. Future work will help to discern between the two proposed different
scenarios.

\section{The X-ray properties from Ultra-luminous X-ray Sources} \label{models}

The spectra of ULXs show a power-law spectral shape in the 3-8\,keV spectral range, together with a high-energy
turn-over at 6-7\,keV (Stobbart et al., 2006; Gladstone et al. 2009; Caballero-Garcia \& Fabian, 2010), and a {\it soft excess} at lower energies (e.g. Kaaret et al., 2006). This {\it soft excess} can be modelled
by emission coming from the inner accretion disc and is characterized by a low inner disc temperature of ${\approx}0.2$\,keV. This
is expected if the black holes in these sources are indeed IMBHs (Miller et al., 2004). {\it Other explanations for the {\it soft excess}
imply a much smaller mass for the BH in these sources, based on the idea that the accretion in the disc is not intrinsically standard,
in contrast to the majority of BHBs (e.g. see Kajava \& Poutanen, 2009)}.

\subsection{The standard accretion disc theory}

The low inner disc temperatures found for some ULXs were initially interpreted as an evidence for the presence of IMBH (Miller et al., 2004).
In the standard disc-black body model (i.e. Multi-Color Disc Blackbody or MCD; Makishima et al., 1986, 2000), which is an approximation 
of the real standard accretion disc theory, the bolometric luminosity from the accretion disc is calculated as:

\begin{equation} \label{eq1}
 L_{\rm bol}=4{\pi}(R_{\rm in}/{\zeta})^{2}{\sigma}(T_{\rm in}/{\kappa})^{4}
\end{equation}

\noindent Here ${\kappa}{\approx}1.7$ (Shimura \& Takahara, 1995) is the ratio of the color temperature to the effective temperature, or ''spectral hardening
factor'', and ${\zeta}$ is a correction factor taking into account the fact that $T_{\rm in}$ occurs at a radius somewhat larger than $R_{\rm in}$
(Kubota et al., 1998 give ${\zeta}=0.412$). However, a recent spectral study of the spectral variability
from a sample of ULXs, including NGC~5408 X--1 (Kajava \& Poutanen, 2009), has shown that the {\it soft excess} (i.e. the disc component fitted in the spectra) from NGC~5408 X--1
does not follow Eq.~1 but $L_{\rm bol}{\propto}T_{\rm in}^{-3.5}$. {\it This in contrast to what is found for many BHBs and might indicate that
the standard accretion disc theory is not a proper interpretation in the case of NGC~5408 X--1. This implies that the hypothesis on which
the IMBH idea is relying (i.e. standard accretion disc theory and the presence of a cold disc) are not valid and
it might indicate that the BH in NGC~5408 X--1 is not an IMBH}.

\subsection{The SLIMULX model} \label{slimulx}

Our analysis is based on the SLIMULX model. SLIMULX is an additive model for
thermal continuum emission at high accretion rates to be used with the
X-ray spectral-fitting tool XSPEC. The model provides spectral
distribution of black-body radiation that is supposed to be emitted from
the surface of a slim accretion disc (Abramowicz et al., 1988). It uses numerical solutions of
radial disc structure equations (S{\c a}dowski et al., 2011) that go beyond the
standard thin disc model (hereafter referred to as SDT) by including advection of matter and heat or
non-Keplerian rotation throughout the disc. Because the geometrical
thickness of the disc can be considerable, the model estimates the
location of the effective photosphere and computes the radiation
transport in Kerr space-time from that place with all relativistic
effects properly calculated.

The SLIMULX model targets luminous sources that accrete at higher rates and 
for which the thin disc approximation becomes invalid. The higher the
accretion rate, the higher departure from the standard (thin disc)
temperature profile there is as the advection of heat becomes more
important and the more is the disc peak emission shifted inwards due to
advection making the flow less radiatively efficient. The radial profile
of cooling flux in the inner disc regions for super-Eddington accretion
rates deviates from standard ${\propto}r^{-0.75}$ dependence
to ${\propto}r^{-0.5}$. Such behaviour is valid for all values of the viscosity
parameter $\alpha$. However, the higher the value of $\alpha$, the
earlier advection affects the disc emission.

The final model spectrum is built as a sum of the local contributions from
all parts of the disc surface. Some (inner) parts of the disc may,
however, be hidden to the observer because they can be effectively
shielded by outer parts of the geometrically thick disc. The importance 
of this effect grows with accretion rate and with observer inclination. In 
the most extreme situation the observer may be completely shielded from
the inner hottest and most energetic parts of the disc being able to see
only relatively cool and more distant part of the disc surface. The local
contributions to the spectrum are in the frame co-moving with the disc
surface modelled as a simple multi-color black-body. The user has a 
manual control over the overall hardening factor.

The present version of SLIMULX model comes in a form of a FITS table
with pre-calculated spectra and a small routine which reads the data and
produces the final spectrum in terms of interpolation in the parameter
space. For five different values of ${\alpha}$-viscosity and three different
values of scale-height modifier, the table contains extensive three
dimensional grids of spectra by varying the spin, luminosity and
inclination. The three parameters are varied within the range of their
limits on different types of logarithmic scales to ensure that
the spectra make roughly equal changes between any two adjacent steps.

\section{Observations and data reduction} \label{analysis}

For the spectral analysis we used only the {\it XMM-Newton}/EPIC pn camera, in order to avoid issues due to cross-calibration effects. Additionally, the EPIC pn camera
has a higher effective area (i.e. double) than each one of the {\it XMM-Newton}/MOS cameras and has sufficient statistics for the spectral fitting. We used only the data
from the ObsID num. 0653380201, since it has the highest number of counts. We applied the standard time and flare filtering (rejecting high-background periods of rate ${\ge}12$\,counts/s).
We filtered the event files, selecting only the best-calibrated events (pattern${\le}4$ for the pn), and rejecting flagged events (flag$=0$). We extracted the flux from a circular 
region on the source centred at the coordinates of the source and radius $49\,$arcsec (the same as used by Gladstone et al., 2009). The background was extracted from a
circular region, not far from the source. 

We built response functions with the {\it Science Analysis System} (SAS) tasks {\tt rmfgen} and {\tt arfgen}. The
background-subtracted spectra was fitted with standard spectral models
in XSPEC 12.9.0 (Arnaud, 1996). All errors quoted in this work are $68\%$ ($1{\sigma}$) confidence. The spectral fits were limited to the 0.3-8\,keV range, in order to minimize
the effects of the background selection. The spectra were rebinned in order to have at least 100 counts for each background-subtracted spectral channel
in order to perform the chi-squared fitting and to avoid oversampling of the
intrinsic energy resolution by a factor larger than 3.

\section{Spectral analysis and results} \label{spec_anal}

As found in previous studies (Caballero-Garcia et al., 2013b), the X-ray spectrum of NGC~5408 X--1 can be modelled by a continuum formed by an absorbed (curved) power-law describing the hard X-ray emission plus 
a soft X-ray emission component from a cold accretion disc (in the SDT). Here we consider a simple (non-curved) absorbed power-law since we restrict the spectral analysis 
to ${\le}8\,$keV. We obtained a poor fit (${\chi}^{2}/{\nu}=255/100$) with residuals in the form of broad emission lines at energies ${\le}1$\,keV. These excesses at low
energies (around 0.6 and 1\,keV) can be attributed to the diffuse emission from the galaxy. To account for them we had to include two {\tt apec} models (see Caballero-Garcia et al., 2013b
for details).

For the accretion disc emission we adopted the relativistic slim disc component {\tt SLIMULX}. This model assumes similar hypothesis to those presented by Poutanen et al. (2007), but calculating
the geodesics of the photons from a slim-disc configuration consistently according to the effects of general relativity (see Sec.~2.2). In this configuration the advection
and obscuration effects cause significant deviations from the X-ray spectra expected in SDT. These effects are {\it strongly inclination dependent and the luminosity can stay at ${\approx}L_{\rm EDD}$ even if mass
accretion rate is ${\gg}1$}. In the case of high accretion rates the disc is strongly radiation pressure dominated and the mass of the BH can not be determined in a straightforward manner as it happens
in the SDT. 

Therefore we fitted the spectrum of NGC~5408 X--1 with the model {\tt TBabs(apec + apec + slimulx + powerlaw)} in XSPEC. We used the Tuebingen-Boulder ISM absorption model ({\tt TBabs} in XSPEC) to account
for the interstellar absorption (${\rm N}_{\rm H}=7{\times}10^{20}\,{\rm cm}^{-2}$ in the direction to NGC~5408; Dickey \& Lockman, 1990). This parameter was set free to vary
in order to account for intrinsic absorption. The distance to NGC~5408 X--1 and the spectral hardening factor ${\kappa}$ were fixed to their canonical values, i.e.
$D=4.8$\,Mpc and ${\kappa}=1.5$. After some trials we found that the best values for the vertical scale of the disc and the spin of the BH were $scaleh=1$
and $a/M=0.99$, respectively. The best fit obtained (${\chi}^{2}/{\nu}=86/95$) gave a BH mass value of $M=5.7{\pm}0.3\,{\rm M}_{\odot}$. We refer to Tab.~1 and Fig.~1
for the values of all the parameters found and the final fitted spectrum, respectively.

\begin{table}[tb]
\begin{center}
  \caption{Results from the spectral analysis.}
  \label{table_spe}
  \begin{tabular}{@{}lc@{}}
  \hline
 \hline
   Spectral parameter                                     &   Value                     \\
 \hline
 \hline
   $N_{\rm H}$\,$({\times}10^{22})\,({\rm cm}^{-2})$      &   $0.143{\pm}0.002$    \\
   $kT_{1}$\,(keV)                                        &   $1.00{\pm}0.02$          \\
   $kT_{2}$\,(keV)                                        &   $0.22{\pm}0.02$          \\
   $M$\,(${\rm M}_{\odot}$)                                     &   $5.7{\pm}0.3$         \\
   $a/M$                                                  &   $0.99$                    \\
   $L_{\rm disc}/L_{\rm EDD}$                             &   $2.9{\pm}0.4$              \\
   ${\theta}_{0}$\,(deg.)                                 &   ${\le}26$         \\
   $D$\,(kpc)                                             &   $4\,800$         \\
   ${\kappa}$                                             &   $1.5$         \\
   $scaleh$                                               &   $1.0$         \\
   ${\Gamma}$                                             &   $3.4{\pm}0.2$           \\
   ${\chi}^{2}/{\nu}$                                     &      $86/95$                \\
 \hline
\hline
\end{tabular}
\end{center}
\end{table}

\newpage

\vspace{10cm}

\includegraphics{fig1a.ps}

\begin{figure}[h]
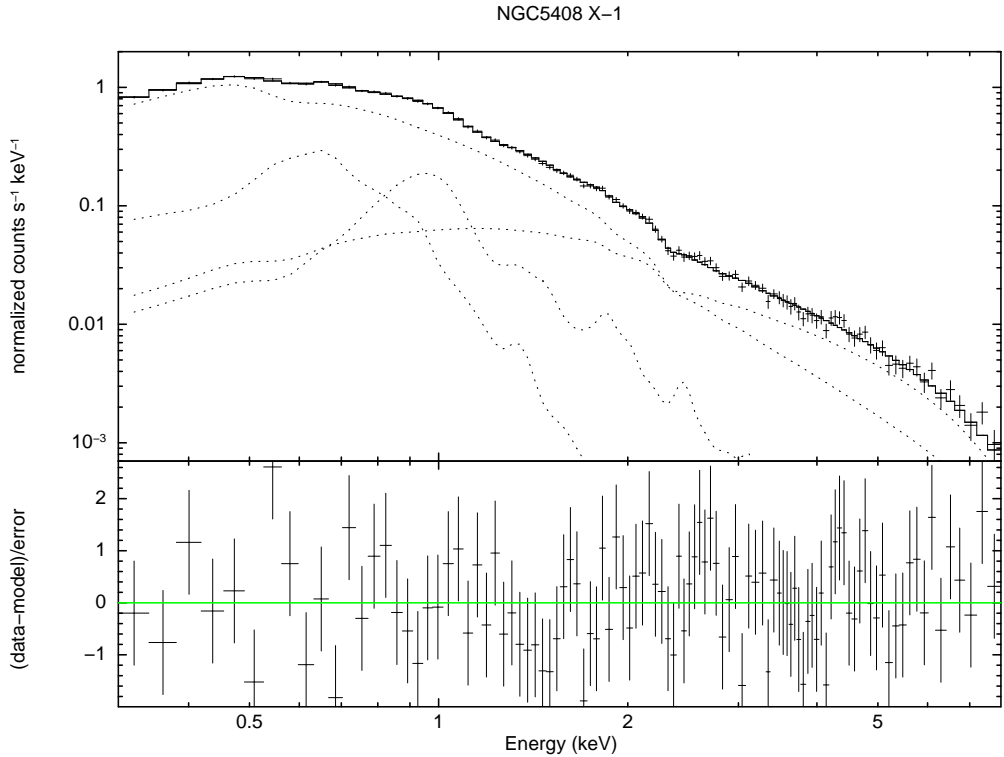

\caption{EPIC-pn {\it XMM-Newton} spectrum (top) and chi-square residuals (bottom) of NGC~5408 X--1 fitted with the spectral model shown in the text. }
\label{fig1}
\end{figure}

\vspace{10cm}

\section{Discussion and conclusions}

In this paper we briefly presented a model ({\tt SLIMULX}) which deals with the X-ray thermal emission from a slim disc around a stellar mass black hole in the super-Eddington 
regime. This model takes into account all the relativistic effects acting on the light in the vicinity of the BH.

We present some preliminary interesting results obtained through the analysis of the X-ray spectrum from NGC~5408 X--1 using this model. The global ($0.3-8$\,keV) X-ray spectrum
is well fitted including this model. We find that the BH is {\it maximally-spinning and the disc is close to face-on} (${\theta}_{0}{\le}26$\,deg). We derive also that
the system is accreting only slightly above the super-Eddington limit ($L_{\rm disc}/L_{\rm EDD}=2.9{\pm}0.4$), giving a {\it mass for the BH of $M=5.7{\pm}0.3\,{\rm M}_{\odot}$, thus not being an IMBH}.

The results obtained are consistent with those previously reported by Poutanen et al. (2007). They proposed
that at high accretion rates an outflow forms within the so-called spherization radius (see Middleton et al., 2015 for an interpretation based on a disc-wind scenario). For a face-on observer the luminosity is high because of geometrical beaming (King et al., 2001). Such an observer has a direct view of the
inner hot accretion disc, which has a peak temperature ${\rm T}_{\rm max}{\approx}1$\,keV in stellar-mass BHs. In this model the {\it soft excess} corresponds to the emission from the spherization radius. Therefore, having 
a stellar-mass BH implies the presence of a much hotter inner accretion disc (i.e. with temperatures higher than the {\it soft excess}), that is 
observed in the spectrum if the inner disc inclination is low. Such a super-Eddington flow implies much lower values for the mass of the BH,
i.e. ${\rm M}{\approx}10\,{\rm M}_{\odot}$, accreting at mildly super-Eddington rates (${\dot M}/{\dot M}_{\rm EDD}{\approx}10$).

\bigskip \noindent {\bf Acknowledgments.}
MCG, MB and MD acknowledge support provided by the European Seventh Frame-work
Programme (FP7/2007-2013) under grant agreement n$^{\circ}$ 312789. SF acknowledge 
support by the Russian Science Foundation (N 14-50-00043).

\end{document}